**Magnetic control of large room-temperature polarization**

**Ashok Kumar, G. L. Sharma, and R. S. Katiyar***

Department of Physics and Institute for Functional Nanomaterials, University of Puerto Rico, San Juan, PR 00931-3343 USA

**R. Pirc and R. Blinc**

Department of Theoretical Physics, Jozef Stefan Institute, Ljubljana 1000, Slovenia

**J. F. Scott***

Department of Earth Sciences, University of Cambridge, Cambridge CB2 3EQ, U. K.

Numerous authors have referred to room-temperature magnetic switching of large electric polarizations as “The Holy Grail” of magnetoelectricity. We report this long-sought effect using a new physical process of coupling between magnetic and ferroelectric relaxor nano-regions. Solid state solutions of PFW [$Pb(Fe_{2/3}W_{1/3})O_3$] and PZT [$Pb(Zr_{0.57}Ti_{0.43})O_3$] are bi-relaxors, with both ferroelectric relaxor qualities and magnetic relaxor phenomena. Near 20%PFW the ferroelectric relaxor state is nearly unstable at room temperature against long-range ferroelectricity. Here we report magnetic switching between the normal ferroelectric state and the ferroelectric relaxor state. This gives both a new room-temperature, single-phase, multiferroic magnetoelectric, $PbZr_{0.46}Ti_{0.34}Fe_{0.13}W_{0.07}O_3$, with polarization, loss (<4%), and resistivity (typically $10^8$ - $10^9$ ohm.cm) equal to or superior to $BiFeO_3$, and also a new and very large magnetoelectric effect: switching not from $+P_r$ to $-P_r$ with applied H, but from $P_r$ to zero with applied H of less than a Tesla. This switching of the polarization occurs not because of a conventional magnetically induced phase transition, but because of dynamic effects:

Increasing H lengthens the relaxation time by x500 from <200 ns to >100 μs, and it couples strongly the polarization relaxation and spin relaxations. The diverging polarization relaxation time accurately fits a modified Vogel-Fulcher Equation in which the freezing temperature $T_f$ is replaced by a critical freezing field $H_f$ that is 0.92 ± 0.07 Tesla. This field dependence and the critical field $H_c$ are derived analytically from the spherical random bond random field (SRBRF) model with no adjustable parameters and an $E^2H^2$ coupling. This device permits 3-state logic ($+P_r$,0,$-P_r$) and a condenser with >5000% magnetic field change in its capacitance; in H=0 the coercive voltage is 1.4 V across 300 nm for $+P_r$ to $-P_r$ switching, and the coercive magnetic field is 0.5 T for +Pr to zero switching.

***Corresponding authors**

---

Email: rkatiyar@uprrp.edu (Prof. Ram S Katiyar) and Email: jsco99@esc.cam.ac.uk (Prof. James F Scott)

At the end of the 19th Century, magnetoelectric effects -- the manipulation of magnetization by electric field and induction of polarization by magnetic field – were suggested by Pierre Curie [1]. In 1957-9 these suggestions were proved for $Cr_2O_3$ theoretically by Dzyaloshinskii and experimentally by Astrov in Moscow. And afterwards in the 1970s careful studies were carried out by many groups[2-4] on various magnetoelectric materials. In the last decade related studies on ferroelectrics that are weakly magnetic[2,3,4,5] and magnets that are weakly ferroelectric[6,7,8] have been done in order to get novel magnetoelectric materials. As far as we know, there are few single-phase materials in nature that posses both ferroelectric and ferromagnetic properties independently[9]. Magnetoelectric multiferroics are at present defined as single-phase materials[5-11] or artificially designed nanostructuctures[10,11]where different ferroic orders such as (anti)ferroelectricity, (anti)ferromagnetism, and ferroelasticity coexist and at least one magnetic and one electric order parameters are coupled with each other. After the fabrication of epitaxial thin films of $BiFeO_3$ (BFO) with room-temperature multiferroic behavior ($P_r$ ~55 μC/cm$^2$ @ 15 kHz, $M_s$ ~ 150 emu/cm$^3$)[12,13], much multiferroics research has centered around this material; yet it is well known that BFO is not an ideal magnetoeloectric: It often shows very high leakage current and applied H does not much affect polarization P.[14,15].

Switching: Recently there has been a serious research effort in many countries to produce electric switching of magnetizations or magnetic switching of polarizations in multiferroic magnetoelectrics. Generally schemes for switching from +M to –M with electric field E have been examined, or conversely from $+P_r$ to $-P_r$ with magnetic field H. $BiFeO_3$ has been the material of choice at room temperature, but in addition over the past several years a useful hint to the materials design of multiferroics has been obtained by the studies on other magnetic ferroelectrics, including manganites. The rare-earth maganite $RMnO_3$ (R= Tb, Dy) have shown strong magnetoelectric (ME) properties near the magnetic transitions[2,3,16]. Unfortunately these groups of materials show magnetic behavior only at cryogenic temperatures (< 50 K). There are several other magnetically frustrated systems that have been identified which also showed gigantic magnetoelectric effects at low temperature. Among these materials $MnWO_4$ showed strong temperature and external magnetic field dependent of its ferroelectric loops. Recently Kundys et al.

observed magnetic field-induced ferroelectric hysteresis loops in $Bi_{0.75}Sr_{0.25}FeO_{3-\delta}$ materials[17] and effects of external magnetic field on the ferroelectric hysteresis loops of $MnWO_4$ single crystals[18]. These two results and those of reference 19 are of our particular interest because they showed similar behavior to that in the present investigation.

In the present scenario a new single-phase material is needed that can exhibit magnetoelectricity (not necessarily linear) at room temperature. In search of such new magnetoelectrics we have taken a different approach: Ignore subtleties of symmetry and examine multirelaxors in which strong $E^2H^2$ coupling exists independent of long-range symmetry. We found a single-phase $Pb(Fe_{0.66}W_{0.33})O_3$(PFW)-$PbTiO_3$ (20:80) solid solution that showed electric dipole order near room temperature with very low loss below its phase transition temperature[19]. Smolenskii et al. discovered various related multiferroics compounds in late 1950s, among which PFW is one of the most promising candidates, having a ferroelectric relaxor transition near 180 K and an antiferromagnetic (AFM) phase transition (343 K) above room temperature[20]. Since lead zirconate titanate $PbZr_xTi_{1-x}O_3$ (PZT) thin films has been extensively studied for potential and practical applications in dynamic and non-volatile ferroelectric random access memories due to their large remanent polarization ($P_r$), small coercive field and Curie temperature above room temperature[21,22], it therefore seemed useful in order to exploit the basic properties of these two materials to make a complex single-phase material using a chemical solution technique (CSD). Table 1 shows the properties of PFW/PZT for 20%, 30%, and 40% PFW. All of these samples are weakly ferromagnetic and ferroelectric at some temperature, but note that increasing amounts of PZT raise the relaxor-to-ferroelectric phase transition from about 150K at 60% to 300K at 80%, so that only the 80% PZT specimens combine the most interesting properties at ambient temperatures. We have carried out extensive measurements on three naturally occurring multiferroics PFW[19, 20] $Pb(Fe_{0.50}Nb_{0.50})O_3$, (PFN) [23 ,24], and $Pb(Fe_{0.50}Ta_{0.50})O_3$ (PFT) [24] with the "universal" ferroelectric PZT, since these three multiferroics by themselves are not good enough to provide both good resistivity, multiferroicity, and magnetoelectric coupling. The PFN and PFT results will be shown elsewhere. Note that it is not necessary that the materials studied be ferromagnetic at room temperature, since the coupling we see, although very large, is of form $E^2H^2$. The magnetic hysteresis loops of samples with 20% PFW, 30% PFW, and 40% PFW are shown in Fig.1. The M-H data display weak ferromagnetism. The induced magnetization for 0.2PFW/0.8PZT [i.e., $PbZr_{0.46}Ti_{0.34}Fe_{0.13}W_{0.07}O_3$] was 5 emu/cc at a high magnetic field of 1.5 T. Due to the experimental limitation of applied magnetic field, we had a maximum of only 1.5 T; up until this external field the M-H curve for 20% PFW shows an unsaturated loop with estimated very low saturation magnetization ($M_s$~ 0.48 emu/cc) and very low coercive field (H ~ 0.12 T), but at 40% the saturation is better and $M_s$ = 4.5 emu/cc. A super-exchange in the disordered regions through $Fe^{+3}$-O-$Fe^{+3}$ is expected to yield antiferromagnetic ordering[19,25].

X-ray diffraction patterns (XRD) were obtained of the $PbZr_{0.46}Ti_{0.34}Fe_{0.13}W_{0.07}O_3$---PZTFW46/34/13/7 thin films deposited on $Pt/TiO_2/SiO_2/Si$ substrates at 400°C and post-annealed from 600°C to 750°C for different Rapid Thermal Annealing (RTA) time (See Supplementary Fig. S1 (a)). As-grown films are amorphous in nature. The XRD analysis indicates that the films grown at 700 °C were single-phase polycrystalline in nature with less than 0.1% impurities (pyrochlore) phases. This complex single-phase perovskite possesses more than six elements, which makes it very unlikely to get a 100% pure phase, but the electrical properties of all the films having less than 5 % pyrochlore (anything annealed above 600$^{o}$C) were excellent, which indicates that these pyrocholore components do not damage the electrical response. A preliminary XRD investigation was carried out with the well-known POWD program[26] which fitted well a tetragonal crystal structure having lattice parameters a =4.0217 Å and c = 4.0525 (standard deviation each of 0.0060). The intensity of the (100), and (110) peaks of PZTFW46/34/13/7 films increases with increase in temperature, illustrating a better crystalline state and enhancement in grain size. In this paper electrical, magnetic and dielectric characterization were carried out for as-grown films at 400°C with further annealing at 700°C for 60 seconds (for more complete crystallization continuously annealed additionally at 650°C for 180 seconds). The surface morphology of the films was investigated by AFM in contact mode over a 5 x 5 μm area and 20 nm heights (See Supplementary Fig. S1(b)) which indicates well-defined grains with an average size 1 μm. Surface topography image of a 5 x 5 μm region revealed average surface roughness of 10.2 nm. The observed bigger grain size and higher surface roughness may be due to growth at high temperature and utilization of the conventional chemical solution deposition process[14].

The dielectric constant and dielectric loss (inset) measured as functions of temperature and frequency are shown in Figure 2. Fig. 2 shows a broad dielectric dispersion from 100 Hz to 1 MHz over a wide range of temperature for higher frequency.

The electric field induced polarization switching (P-E) behavior was studied by Sawyer-Tower measurements at 60 Hz. The films exhibit well saturated hysteresis loops with remanent polarization ($P_r$) and the coercive field ($E_c$) of about 70 $\mu C/cm^2$ and 48 kV/cm (1.4 V across 300 nm) respectively for 350 kV/cm maximum external electric field (See Supplementary Fig. S2). We did not observe much change in the coercive field with increase in applied electrical field. The observed $P_r$ value is higher or at par with the reported value of polycrystalline PZT thin films[27]. Enhancement in remanent polarization may be attributed to the relaxed local strain (due to bigger grains) and higher nucleation of grains and compositional disorder due to the solid solution of ferroelectric and relaxor. In order to check the real polarization behavior of PZTFW46/34/13/7 films and its utility in memory devices, we performed fatigue tests. Fatigue behavior of PZTFW46/34/13/7 thin films was carried out with 100 kHz bipolar square waves in the presence of 350 kV/cm external electric field (See Supplementary Fig. S2). There was a modest polarization loss (<12%). Twelve percent decay in fatigue of PZTFW46/34/13/7 thin films on a platinized silicon substrate after $10^9$ cycles is much better than the earlier reports for PZT,[28] indicating its suitability for memory device applications.

Three-state logic and switching at room temperature is shown in Fig.3. $+P_r$ to $-P_r$ switching is observed at $E_c$ = 48 kV/cm (1.4 V across 300 nm); and magnetic switching from +Pr to zero is shown at 0.5 Tesla.

The polarization response at 0.5T is shown in the insert in Fig. 3; it is that of a leaky linear dielectric. The reason for this behavior is shown clearly in Fig. 4, where we have graphed the dielectric constant versus frequency at different magnetic fields. Note that, for example, the dielectric constant at H = 0.80T is 2400 for f = 1 kHz but 390 at f = 100 kHz. This is simply due to the field dependence of the relaxation time, and the fact that as H increases the size (correlation length) of spin clusters in this magnetic relaxor increases and hence their fluctuation time decreases (in an extrapolated infinite field H, the system would have long-range ferromagnetic order and the relaxation frequency would go to zero). The fact that this strongly influences their dielectric susceptibility requires strong coupling between spin fluctuations and polarization fluctuations. In a separate paper we relate these observations to the recent discovery of multiferroic relaxors by Levstik et al.[29] and to the model of magnetic-relaxor/ferroelectric-relaxor coupling of Shvartsman, Kleemann et al.[30]. In the theory section below we show that these results can be derived analytically from the spherical random-bond random-field (SRBRF) model and that the key term is biquadratic $E^2H^2$. Fig.4 shows that the peak of the dielectric loss at H = 0.85T is at f = ca. 2-3 kHz. Note that this agrees with the inflection frequency of 2-3 kHz estimated from real part of permittivity of Fig. 4, suggesting a magnetic field-induced Debye relaxation.

The supplementary data (figure S3) show the magnetic field dependent Cole-Cole plot of dielectric spectra, fitted to a stretched exponential. The relaxation frequencies obtained from the characteristic peak in dielectric loss spectra from Fig.5 and/or from the stretched exponential are plotted in Fig. S4. These characteristic frequencies can be accurately fitted either of two Vogel-Fulcher-type equations:

$$f = f_0 \exp\left[\frac{-E_a}{\mu_B\left(H_f - H_R\right)}\right] \quad \text{(1a)}$$

or

$$f = f_0 \exp\left(-\frac{U_2}{H_f^2 - H_R^2}\right) \quad \text{(1b)}$$

where $H_f$ is a freezing field and replaces the analogous term in $T_f$ (freezing temperature) in the usual Vogel-Fulcher Equation; $\mu_B$ is the Bohr magneton; and $H_R$, the relaxation field. Although the diverging polarization relaxation time accurately fits this modified Vogel-Fulcher Equation, Eq.1a arises from an $E^2H$ term in the free energy which can be ruled out by the independence of the polarization data upon the sign of applied field H. Eq.1b arises from an $E^2H^2$ term in the free energy and is independent of the sign of H. The fitted data provide freezing field $H_c = 0.92 \pm 0.07$ Tesla and characteristic frequency $f_0 = 40 \pm 1$ MHz. The coupling of order parameters in relaxors is addressed in refs [32,33] and is detailed below, using the spherical random-bond random field model. The coupling constant is taken as real; the general description of coupled oscillators permits either real or imaginary coupling, with the latter implying decay into the same final state[31]. Tokunaga et al. reported for $Dy_xBi_{1-x}FeO_3$ that if the applied magnetic field (H) is greater or equal to the reorientation of Fe spins ($H_{re}^{Fe}$), it produces a weak ferromagnetic component along the c-axis of the crystal and is able to generate or flop the ferroelectric polarization[16]. That is a kind of magnetic phase transition. But what we invoke in the present case is somewhat more complex and not strictly phase transition at all, because like all relaxors, it is frequency-dependent; it is the decrease in relaxation time below the measuring probe frequency, brought on by spin alignment and increased spin correlation length in increased external magnetic fields.

Theory:

We now discuss the origin of the magnetic-field modified Vogel-Fulcher (VF) relaxation rate observed in 0.2PFW/0.8PZT. It has been suggested earlier[33] that the basic relaxation mechanism in relaxor ferroelectrics is growth and percolation of polar nanoregions (PNRs). The main assumption is that the local electric field inside the polarization cloud of a PNR falls off with distance as $\sim 1/r^3$. The potential energy of an induced dipole at $r$ is balanced against the thermal fluctuation energy $\sim kT$, thus yielding the correlation volume $v_c \sim 1/T$. The volume fraction of PNRs then grows as $\sim 1/T$ until the percolation limit $T_p$ is reached, corresponding to the VF temperature $T_0$.

The local electric field in a PNR in the presence of magnetoelectric (ME) coupling will be derived from the Landau-type free energy of a multiferroic system

$$F_0(P,M) = \frac{1}{2}\chi_{m,i}^{-1}P_i^2 + \frac{1}{2}\chi_{m,j}^{-1}M_j^2 + \frac{1}{4}b_e P^4 + \frac{1}{4}b_m M^4 + \cdots - E_i P_i - H_j M_j . \qquad (2)$$

The dielectric susceptibility tensor is diagonal, $\chi_{e,ij} = \delta_{ij}\chi_{e,i}$, representing the response of a disordered ferroelectric close to the boundary between the ferroelectric and relaxor phase. An explicit expression for $\chi_{e,i}$ can be obtained from the SRBRF model of relaxor ferroelectrics[34]. Similarly, the magnetic susceptibility $\chi_{m,j}$ for a weakly ferromagnetic subsystem is defined in S.I. units[35] as $M_i = \chi_{m,ij}H_j$. The anharmonic coefficients $b_e$, $b_m$ formally ensure thermodynamic stability. No direct ME coupling terms appear in Eq. (2) since experimental evidence of them is missing. However, an indirect ME effect[8] will be induced via electrostrictive and magnetostrictive strains, $u_k = Q_{e,ki}P_i^2$ and $u_l = Q_{m,jl}M_j^2$, with electro- and magnetostriction coefficients $Q_{e,ki}$ and $Q_{m,lj}$, respectively. These are related to the inverse susceptibility tensors by the Maxwell relations[36] $Q_{ki} = -(1/2)(\partial\chi_i^{-1}/\partial X_k)_T$, where the stresses $X_k$ are related to strains through the elastic constants $X_k = C_{kl} u_l$. The Voigt notation and summation convention are implied as appropriate.

Expanding the inverse susceptibilities in Eq. (2) to linear order in $X_k$, applying the Maxwell relations, and adding the elastic energy $(1/2) C_{kl}^{-1} X_k X_l$, we obtain after minimizing the free energy a new fourth-order ME term

$$F_1(P,M) = -\frac{1}{2} \lambda_{ij} P_i^2 M_j^2, \qquad (3)$$

where the ME coupling coefficient is $\lambda_{ij} = 2 C_{kl} Q_{e,ki} Q_{m,lj}$.

The equilibrium condition $\partial(F_0 + F_1)/\partial P_i = 0$ yields the electric field $E_i = \chi_{e,i}^{-1} P_i (1 - \chi_{e,i} \lambda_{ij} M_j^2)$ to linear order in *P*. It immediately follows that the local electric field within each PNR similarly acquires an additional contribution proportional to $M_j^2 = \chi_{mj}^2 H_j^2$. This contribution now appears at each step in the derivation of the VF equation[33].

For any direction of $H_j$ all orientations of PNR polarizations $P_i$ are allowed. Moreover, in view of random isotropy the average volume fraction of PNRs is independent of the direction of *H*. Performing a linear average of $\lambda_{ij}$ over *i* and *j,* and introducing hydrostatic coefficients $Q_h = Q_{11} + Q_{12} + Q_{13}$ we obtain the averaged coupling constant $\bar{\lambda} = 2 C_h Q_{e,h} Q_{m,h}$, where $C_h = \sum_{kl=1}^{3} C_{kl}$ is the bulk modulus.

The VF temperature $T_0$ thus becomes a field dependent quantity $T_0(H) = T_0(1 - \chi_e \chi_m^2 \bar{\lambda} H^2)$. The VF relaxation time diverges on the line of percolation critical points $T = T_0(H)$ in the *T,H* plane. The sign of $\bar{\lambda}$ is determined by the signs of $Q_{e,h}$ and $Q_{m,h}$, which can be either positive or negative. If $\bar{\lambda} < 0$, the PNRs will freeze at a VF temperature higher than its zero-field value $T_0$. This seems to be the case in 0.2PFW/0.8PZT. The VF relaxation rate at fixed temperature can be written as a function of the magnetic field

$$f = f_0 \exp\left( -\frac{U_2}{H_c^2 - H^2} \right), \qquad \text{for } H < H_c, \tag{4}$$

in agreement with the empirical result (1b). The barrier height $U_2 = H_0^2 U / kT_0$ is given in terms of the zero-field VF parameters $U / kT_0$ and a scaling field $H_0^2 = 1/(\chi_e \chi_m^2 |\bar{\lambda}|)$. The critical field is given by $H_c^2 = H_0^2 (T - T_0)/T_0$ or

$$H_c^2 = \frac{T - T_0}{T_0} \frac{1}{2\chi_e \chi_m^2 C_h |Q_{e,h} Q_{m,h}|}. \tag{5}$$

Thus $H_c$ can be expressed in terms of independently measurable physical parameters of the system.

Although the parameters occurring in this expression are generally unknown for 0.2PFW/0.8PZT, using averages of known values of electrostriction and magnetostriction in the quasi-isotropic approximation $Q_{eff} = Q_{11} + 2Q_{12}$ for other perovskite oxides[36] yields an estimate of $H_c^2 = 0.3 \pm 0.2$ $T^2$, i.e., $H_c = 0.5 \pm 0.2$ Tesla, giving factor of two agreement with the experimental $H_c = 0.92$ T. Our theory does not yield the sign of the coupling constant $\bar{\lambda}$, which experimentally is negative (applied $H$ turns ferroelectricity into relaxor behavior).

In conclusion, this manuscript reports the successful synthesis of a novel single-phase complex perovskite room-temperature magnetoelectric multiferroic. A narrow temperature window was found to get the desired phase. High dielectric constant, low dielectric loss, highly frequency-dispersive susceptibility, more than 50% temperature dependent dielectric tunability, and above-room-temperature dielectric maxima were observed in PZTFW46/34/13/7 thin films. They show a high ferroelectric polarization of 70 $\mu C/cm^2$ and low coercive field of 48 kV/cm with 12% fatigue in polarization after $10^9$ cycles, thus suggesting a potential candidate for memory applications. At room temperature a weak ferromagnetic/antiferromagnetic M-H hysteresis was detected which suggests an extra degree of freedom of Fe ions in complex octahedra, tilting of $Fe^{+3}$-O-$Fe^{+3}$ linkage, and charge ordering with nearest neighbors. The ferroelectric hysteresis flops under the application of external magnetic field (> 0.50 T), revealing a strong magnetoelectric coupling in the present system, but this is not due to a conventional magnetically-induced phase transition (instead, due to a field-dependent spin relaxation time). This new magtnetoelectric effect, switching from $+P_r$ to zero with applied magnetic field, is 1000x greater than in rare-earth manganites and occurs at room temperature, suggesting a variety of new magnetoeloectric devices, including three-state logic elements.

**Methods.**

Ferroelectric PZTFW46/34/13/7 thin films were deposited on $Pt/Ti/SiO_2/Si(100)$ substrates using chemical solution deposition (CSD). The high purity (> 99.9 %) precursor materials (i.e., lead acetate trihydrate, iron 2-4 pentanedieonate, titanium isopropoxide, zirconium isopropoxide and tungsten isopropoxide) were brought from Alfa Aesar and used for making stock solutions. For good quality films a PZTFW46/34/13/7 solution of 0.2-0.3 molar concentration was spin-coated at 3000 rpm for 30 seconds and pyrolyzed at $400^0$C for 2 minutes. This process was repeated several times in order to get the desired film thickness. Rapid thermal annealing (RTA) was performed at different temperatures from 600-$750^0$C for different time intervals ~ 60-300 seconds to get the desired phase and high density. DC sputtering was carried out for depositing the Pt top electrode of 3.1 x $10^{-4}$ $cm^2$ area using a shadow mask. The dielectric properties in the frequency range of 100 Hz to 1 MHz were studied using an impedance analyzer HP4294A (from Agilent Technology Inc.) over a wide range of temperature attached to a temperature controlled probe station (MMR Technology). Magnetic properties were investigated using a vibrating sample magnetometer (VSM) (Lakeshore model 7400). Room-temperature magneto-electric effects were measured using a combination of impedance analyzer, ferroelectric tester (Radiant technology), VSM and home-made probe station. The polarization versus electric field (P-E) hysteresis loop of the capacitor was measured using the RT 6000HVS ferroelectric tester (Radiant Technology) operating in the virtual ground mode. Fatigue behavior was characterized by applying rectangular pulses at 100 kHz and 350 kV/cm.

**Table 1**

| Characteristic | 20%PFW/80%PZT | 30%PFW/70%PZT | 40%PFW/60%PZT |
|---|---|---|---|
| structure | Tetragonal (nm)<br>a = 0.4022±0.0006<br>c = 0.4053±0.0006 | Cubic<br>a = 0.4012±0.0003 | Cubic<br>a = 0.04007±0.003 |
| Electrical polarization | $P_r$ = 70 μC/cm$^2$ at 295K | $P_r$ = 60 μC/cm$^2$ at 200K | $P_r$ = 40 μC/cm$^2$ at 150K |
| Magnetization | $M_r$ = 0.48 emu/cc at 295K | $M_r$ = 2.29 emu/cc at 295K | $M_r$ = 4.53 emu/cc at 295K |
| Dielectric diffusiveness coefficient g | 1.78 ± 0.05 (intermediate ferroelectric/relaxor) | 2.00±0.10 relaxor | 2.00±0.05 relaxor |
| Dielectric constant ε(295K) | 2300 | 1700 | 1200 |
| Dielectric loss δ(295K) | 0.03 | 0.02 | 0.06 |

## Figure Captions

**Table I**: Characteristics of lead iron tungstate – lead zirconate titanate solid solutions.

**Fig. 1** Magnetic hysteresis at room temperature in PFW/PZT samples for (a) 0.2PFW, (b) 0.3PFW, and (c) 0.4PFW.

**Fig. 2** Temperature dependent dielectric constant and loss tangent (inset) over wide range of frequencies (1 kHz to 1MHz): A step-type temperature dependent dielectric constant was observed with maximum value around 400 K, above which it showed constant behavior.

**Fig.3** Three-state logic switching (+$P_r$, 0, -$P_r$) in PFW/PZT: P-E hysteresis under the application of external magnetic field from 0 to 0.5 T changes polarization from $|P_r|$ to zero; application of 1.4V across 300 nm changes $+P_r$ to $-P_r$. Insert is the P = 0 relaxor state on expanded scale, showing a linear lossy dielectric.

**Fig.4** Real and imaginary part of dielectric permittivity (arrows show the direction of real and imaginary part) as function of frequencies under the application of external magnetic field from 0 to 0.85 T.

**Fig. 5** Vogel-Fulcher-type fitting of relaxation frequency; solid curve is a least squares fit to Eqs. 1b.

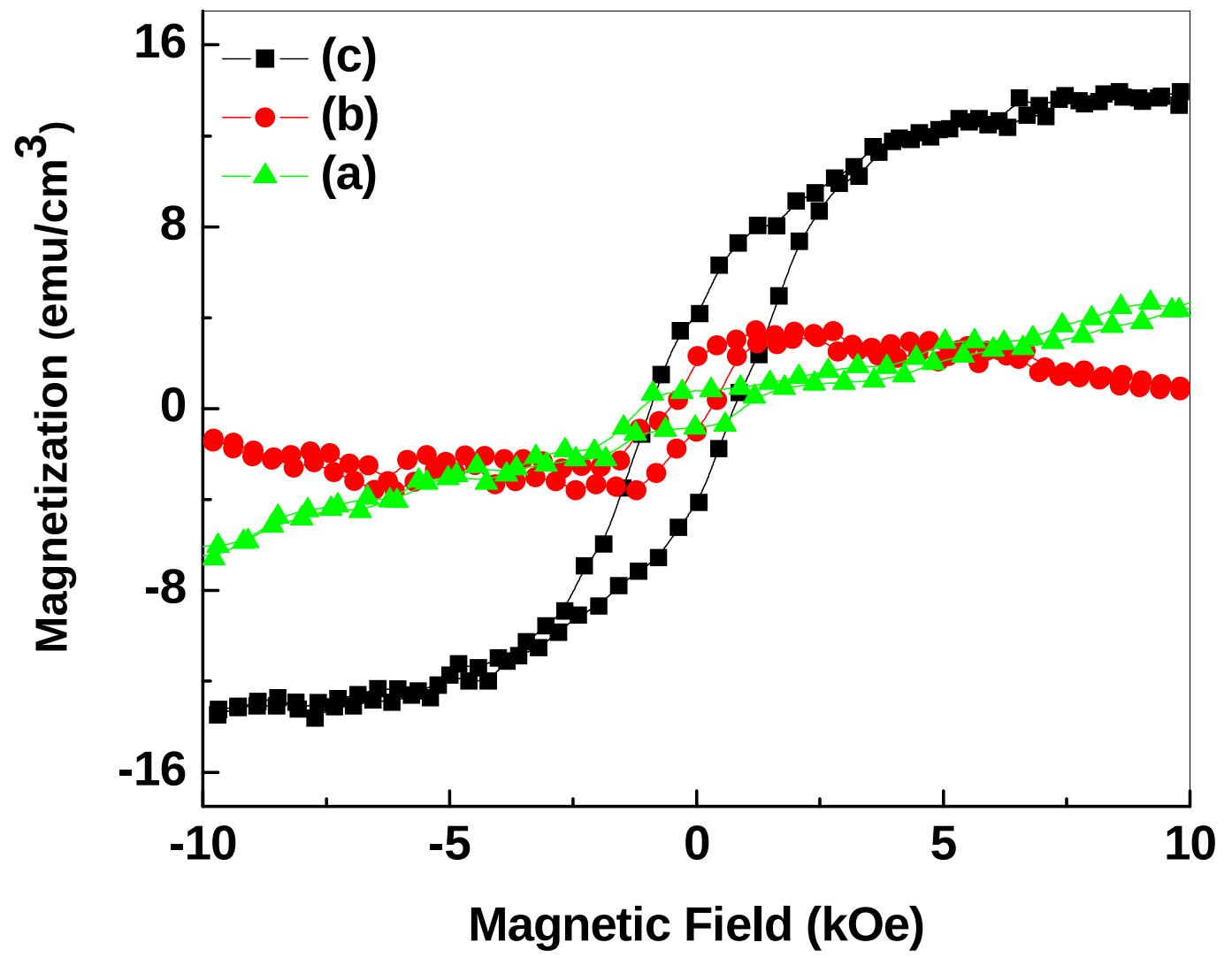

Figure 1

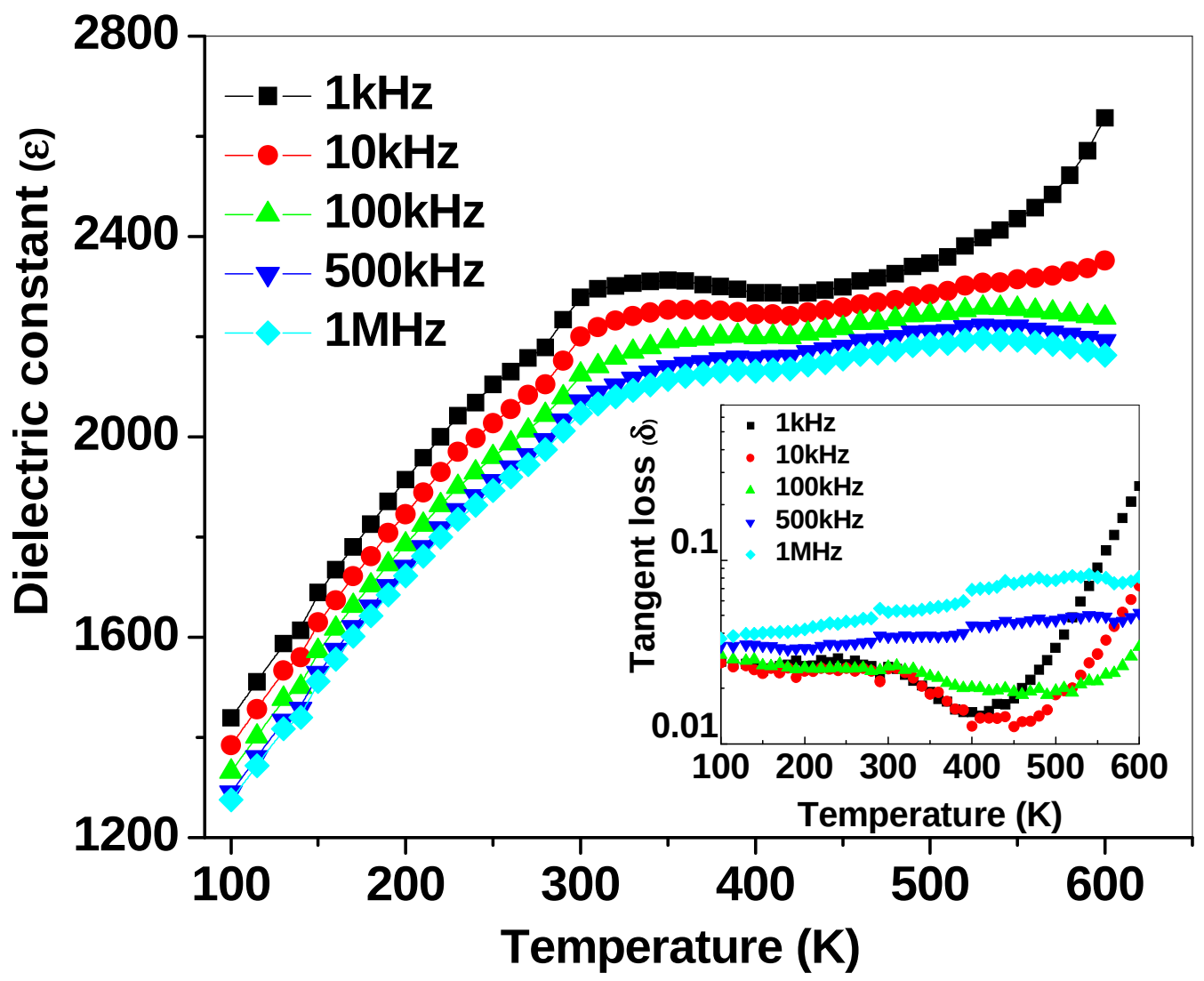

Figure 2

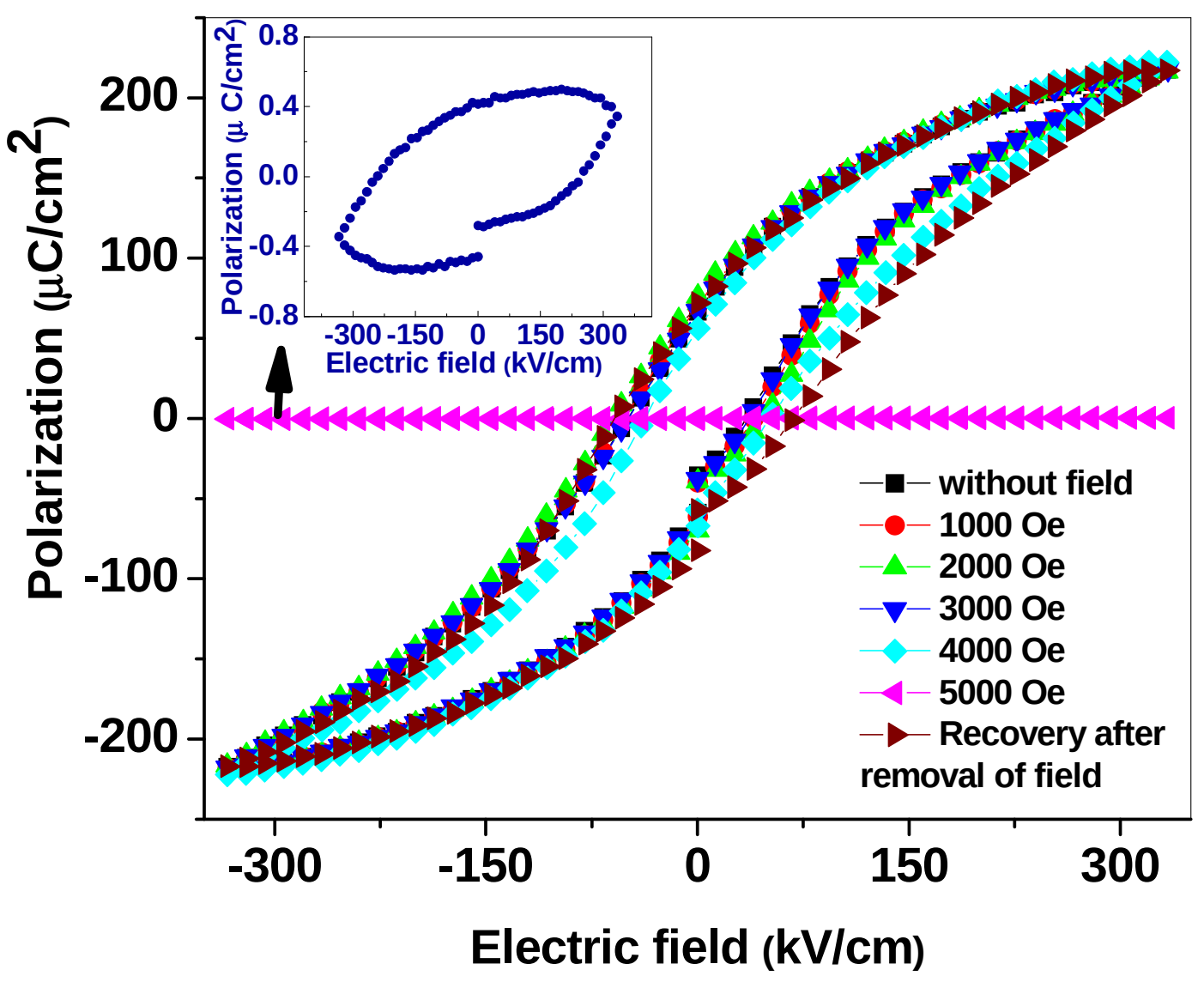

Figure 3

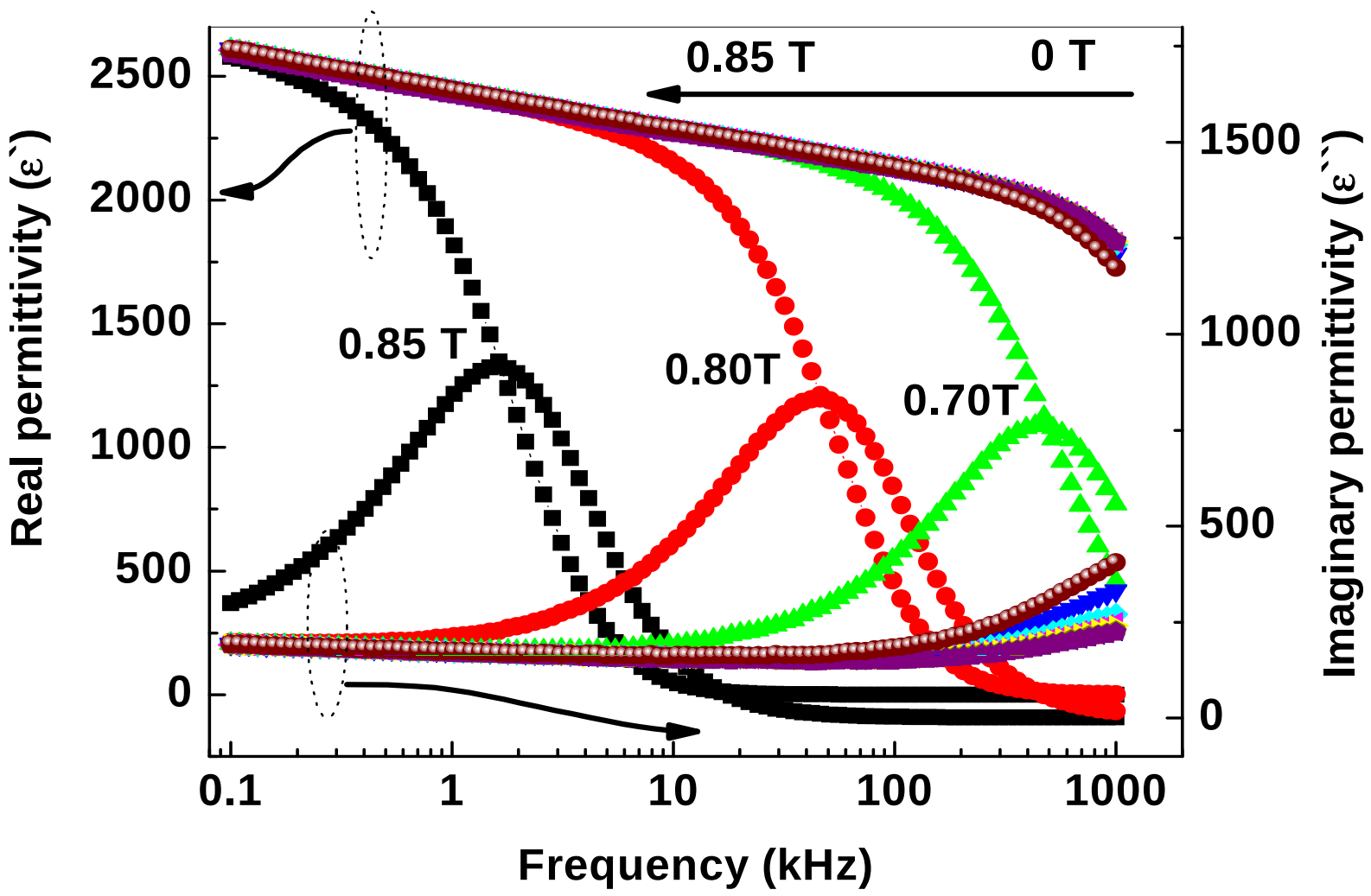

Figure 4

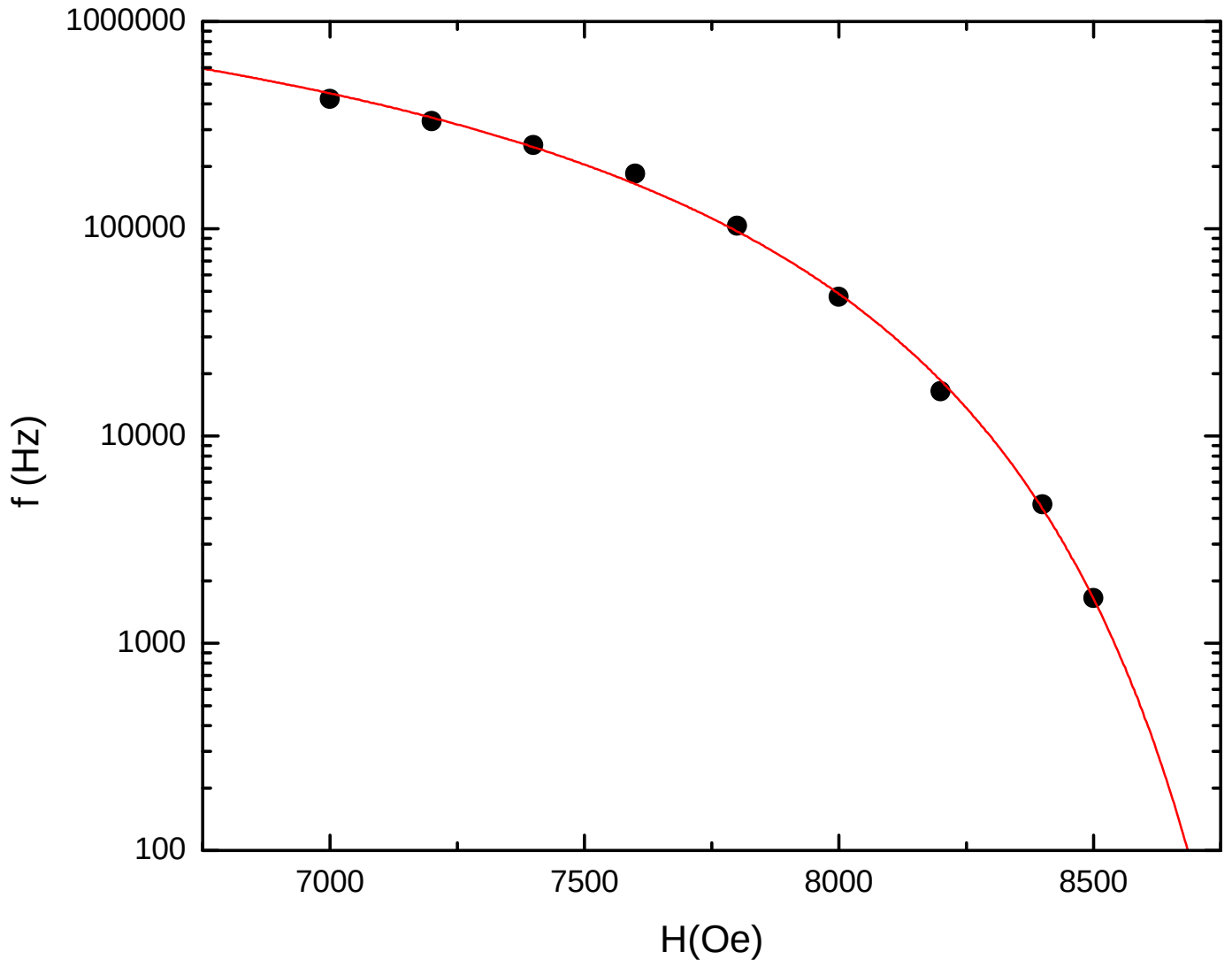

Figure 5